\RequirePackage{fix-cm}
\documentclass[smallextended]{svjour3}       
\smartqed  
\usepackage{graphicx}
\usepackage{amssymb}
\usepackage{amsmath}
\usepackage{cite}
\usepackage[ruled,linesnumbered]{algorithm2e}
\usepackage[justification=centering]{caption}
\begin{document}

\title{High Throughput and Low Cost LDPC Reconciliation for Quantum Key Distribution
}


\author{Haokun Mao   \and
        Qiong Li     \and
        Qi Han       \and
        Hong Guo  
}


\institute{Haokun Mao \and Qiong Li \and Qi Han  \at
              Information Countermeasure Technique Institute, School of
              Computer Science and Technology, Harbin Institute of Technology, Harbin, 150080, China \\   
            \and
            Hong Guo  \at
            State Key Laboratory of Advanced Optical Communication Systems and Networks, and Institute of Quantum Electronics, School
            of Electronics Engineering and Computer Science, and Center for Quantum Information Technology, Peking University, Beijing, 100871, China
            \and
            Qiong Li \email{qiongli@hit.edu.cn} \\
}

\date{Received: date / Accepted: date}

\maketitle

\begin{abstract}
Reconciliation is a crucial procedure in post-processing of Quantum Key Distribution (QKD), which is used for correcting the error bits in sifted key strings. Although most studies about reconciliation of QKD focus on how to improve the efficiency, throughput optimizations have become the highlight in high-speed QKD systems. Many researchers adpot high cost GPU implementations to improve the throughput. In this paper, an alternative high throughput and efficiency solution implemented in low cost CPU is proposed. The main contribution of the research is the design of a quantized LDPC decoder including improved RCBP-based check node processing and saturation-oriented variable node processing. Experiment results show that the throughput up to 60Mbps is achieved using the bi-directional approach with reconciliation efficiency approaching to 1.1, which is the optimal combination of throughput and efficiency in Discrete-Variable QKD (DV-QKD). Meanwhile, the performance remains stable when Quantum Bit Error Rate (QBER) varies from 1\% to 8\%.
\end{abstract}

\section{Introduction}
\label{sec:intro}
\setlength{\parskip}{0.5\baselineskip}
Quantum Key Distribution (QKD) is a promising technique to generate and distribute secure keys with unconditional security for the remote two parties, so called Alice and Bob\cite{Bennett2014Quantum}. To obtain a string of secure key using QKD, two consecutive phases are involved: quantum phase and classical post-processing phase\cite{Gisin2001Quantum}. 
In quantum phase, raw keys are obtained by transmitting and detecting quantum signals via an untrusted quantum channel. Due to the physical noises or the presence of an eavesdropper Eve, raw keys of two parties are weakly correlated and partially secure\cite{RENATO2008SECURITY}. Thus, a distilling phase called post-processing is needed. The main task of post-processing is to convert imperfect raw key strings to consistent secure key pairs via an authenticated classical channel. 
To accomplish this task, a series of post-processing operations have to be performed including sifting, error estimation, reconciliation, verification, privacy amplification and authentication\cite{Walenta2014A}. In this research, we focus on the reconciliation which is often the bottleneck of a high speed QKD system limiting the secure key rate\cite{Dixon2014High}. The main issue of the reconciliation module is to maximize the secure key rate through increased reconciliation efficiency and throughput. Reconciliation efficiency is the parameter that indicates the amount of information leakage. During reconciliation, some information have to be exchanged to correct errors over public channel. The exchanged data may be intercepted by Eve and therefore some information may leak. Throughput determines the amount
of data that can be processed by the reconciliation module. The sifted keys
that cannot be processed or corrected have to be discarded, resulting a decrease
in secure key rate. Unlike reconciliation efficiency, the throughput has not attracted many attentions. However, it has been reported that only an optimal combination of these two parameters: efficiency and throughput, can maximize the secure key rate of a practical QKD system. Especially, throughput plays more important role in high speed QKD systems. 
 
In general, reconciliation protocols can be divided into two categories: interactive and forward error correction code based\cite{Li2014Study}. Cascade is the most widely used interactive reconciliation protocol for its simplicity and relatively high efficiency\cite{Cascade_brassard1993secret,Cascade_sugimoto2000study,Cascade_Nakassis2004Expeditious, Cascade_Yan2008Information, Cascade_Pedersen2013High,Cascade_Pacher2015An}. However, lots of interactions are required in Cascade protocol leading to a degradation in throughput\cite{Li2014Study}. Indeed, nearly all interactive reconciliation protocols are sensitive to latency in the communication channel, which may be unpredictable in practical QKD systems\cite{Dixon2014High}. 
To overcome such shortcomings, forward error correction codes, such as Polar and LDPC, based reconciliation protocols have been proposed. Polar code was proposed by Arikan in 2009\cite{Polar_Arikan2008Channel} and first used in post-processing of QKD by Jouguet in 2013\cite{Polar_jouguet2014high}. The throughput up to 10.9Mbps was achieved when QBER equals to 2\%. However, the Frame Error Rate (FER) was high leading to a limitation of secure key rate. In 2018, Successive Cancellation List (SCL) decoding algorithm was introduced into Polar reconciliation by YAN Shiling \cite{Polar_Yan2018An} resulting in a higher secret key rate. But the efficiency of Polar reconciliation still lags behind LDPC reconciliation with the same frame length. Furthermore, the throughput significantly decreases when the number of paths L increases. 
Nowadays, LDPC reconciliation is adopted in most current high speed QKD systems for its advantages of high efficiency even using short frame length and inherent parallelism\cite{Walenta2014A,QKDSYSTEM_Z201810}. The original LDPC reconciliation is very sensitive to the Quantum Bit Error Rate (QBER), which severely limits its application. In order to achieve high reconciliation efficiency over a wide range of QBERs, rate-adaptive reconciliation protocol was proposed\cite{LDPC_David2011Information}. In 2014, the bi-directional rate-adaptive LDPC reconciliations were implemented in both CPU and GPU\cite{Dixon2014High}, achieving average throughput up to 5Mbps and 15Mbps with high reconciliation efficiency respectively. To further improve reconciliation efficiency, blind reconciliation protocol was proposed\cite{LDPC_Martinez2012Blind}. However, the required round of interaction is more than that in rate-adaptive protocol, which affects the performance gain of throughput. In 2017, E. O. Kiktenko suggested an improved blind reconciliation protocol\cite{LDPC_Kiktenko2017Symmetric}. Via introducing symmetry in operations of two parties and consideration on unsuccessful decoding results, the protocol gains a significant performance increase in both efficiency and interactivity. However, the bi-directional reconciliation strategy cannot be used in the symmetric blind protocol\cite{LDPC_Kiktenko2017Symmetric}, which limits the achievable throughput. 

LDPC reconciliation can be implemented using either hardware (FPGA, ASIC) or software (CPU, GPU). In this research, we focus on software implementations. To implement LDPC reconciliation in software, GPU is commonly used due to its massive parallel computing power\cite{Dixon2014High,GPU_Wang2018High,GPU_Milicevic2018Quasi}. Though the throughput and efficiency can be very high in GPU implementations\cite{Dixon2014High}, the disadvantages of the GPU scheme are obvious too. A GPU device is not able to work alone, but has to work with its host CPU. Moreover, the relatively large volume and high power consumption of a GPU device make it hard to integrate. So the GPU scheme is not well suitable for a compact QKD solution. Comparing to the GPU scheme, the advantage of a CPU scheme is low cost and the disadvantage is much less computing resources. According to the Ref\cite{Dixon2014High}, the throughput of their CPU scheme is only one third of the GPU scheme. How to implement the high performance reconciliation using a low cost CPU scheme has not been well discussed in existing literatures so far. 

Though the amount of computation resources of a CPU device is not as plentiful as a GPU, the modern CPU offers some good features that can be exploited to achieve high performance as well. These features include Single Instruction Multiple Data (SIMD) instruction extensions, multi-core processing module and multi-level caches\cite{CPU_Gal2016High}. SIMD instruction extensions improve the performance of an application by operating on multiple data elements in one instruction instead of processing the data individually. Using a multi-core processing module, each autonomous core can execute the same program with different inputs simultaneously in order to accomplish the computation fast. Multi-level caches can decrease the impact of the mass memory latency which is often the performance bottleneck. 

In this research, a high throughput and low cost software LDPC reconciliation implemented in CPU is proposed. The main contributions of our work are as follows. A quantized RCBP-based LDPC decoder is proposed, which achieves high efficiency using fixed-point representation. Moreover, the average throughput of 57.60Mbps and the average efficiency of 1.108 within a wide range of QBER is achieved using an i7-6700HQ CPU, by taking good advantage of the features offered by modern CPUs. The average speedup factors of $\times$3.8 and $\times$11.3 are achieved by comparison with the latest GPU and CPU works in DV-QKD\cite{Dixon2014High}. It is worth to mention that the performance improvement comes from our algorithm and implementation optimizations rather than the performance boost of hardware. 

The rest of the paper is organized as follows. The LDPC decoding algorithms are introduced in Section 2. In Section 3, the proposed quantized RCBP-based LDPC decoder is detailed. The experimental results and analysis are presented in Section 4. Finally, a brief conclusion is provided in Section 5.  

\section{LDPC decoding algorithm}
\label{sec:ralated}
LDPC code is a kind of linear block code with error correction capability close to Shannon limit. It is originally proposed by Robert Gallager in 1962\cite{LDPC_gallager1962low} and is defined by a sparse parity check matrix $H$ of size $m \times n$. The code rate is defined as $R=1-m/n$, denoting the ratio of information bits. A LDPC code can also be described by a bipartite graph with $m$ check nodes (CNs) and $n$ variable nodes (VNs) corresponding to the rows and columns of $H$ respectively. The matrix element in $H$ can be binary or non-binary alphabets\cite{ryan2009channel}. In this research, only binary LDPC codes are considered for the sake of simplicity. An edge connects $\rm{CN_i}$ and $\rm{VN_j}$ in the bipartite graph if $H_{ij}=1$. The degree of check node $\rm{CN_i}$ named $\rm{deg{CN}_i}$ is defined by the number of ones in the $i^{th}$ row of $H$. Similarly, the number of ones in the $j^{th}$ column is the degree of variable node $\rm{VN_j}$ named $\rm{deg{VN}_j}$.

The iterative Sum-Product Algorithm (SPA)\cite{LDPC_Mackay1999Good} is usually applied to decode LDPC codes for its high decoding efficiency. It consists of two main steps: check node processing and variable node processing. During the decoding process, messages are exchanged between check nodes and variable nodes along the bipartite graph edges. Such message updating process will repeat until the stopping criterion is satisfied. In practice, the sequence of message updating usually influences the convergence of the decoding algorithm. In this research, we focus on layered scheduling\cite{LDPC_Hocevar2004} rather than original flooding scheduling because it enables the decoding convergence to speed-up by a factor of two.

The sign and magnitude of updated output messages in check node processing using layered scheduling can be calculated as Equations (\ref{eq:1}) and (\ref{eq:2}), where $L_{ij}^t$ is the message from $\rm{CN_i}$ to $\rm{VN_j}$ in the $t^{th}$ iteration; $L_{ji}^t$ is the message from $\rm{VN_j}$ to $\rm{CN_i}$ in the $t^{th}$ iteration; $\Psi(i)$ is the set of VNs connected to $\rm{CN_i}$; $\Psi(i)/j$ is the set $\Psi(i)$ excludes $\rm{VN_j}$.
\begin{equation}\label{eq:1}
sign(L_{ij}^t) = \prod\limits_{j' \in \Psi (i)/j}{sign(L_{j'i}^t)}
\end{equation}
\begin{equation}\label{eq:2}
\left| {L_{ij}^t} \right| = 2{\tanh ^{ - 1}}\left[ {\prod\limits_{j' \in \Psi (i)/j} {\tanh \frac{{\left| {L_{j'i}^t} \right|}}{2}} } \right]
\end{equation}

Equation (\ref{eq:2}) can be formulated in two other ways\cite{ryan2009channel} which are presented in Equations (\ref{eq:3}) and (\ref{eq:4}). Mathematically there is no difference between all these three equations, but each equation has its own advantage in practice. For instance, the function $\phi (x)$ used in Equation (\ref{eq:3}) is an invertible function, which is defined in Equation (\ref{eq:5}). By using $\phi (x)$, the multiplication and division operations presented in Equation (\ref{eq:2}) are removed. Furthermore, the function $\phi (x)$ can be implemented by using a look-up table which can be replaced by piecewise linear approximation as discussed in\cite{Decode_Jones2004Approximate, Jones2007Functions}. Although implementation of the $\phi (x)$ function by use of a look-up table is sufficient for some software simulations, but due to dynamic-range issues, it is a difficult function to approximate even with a large table. Unlike the Equation (\ref{eq:3}), the input and output range keeps stable in Equation (\ref{eq:4}). The The “box-plus” operator used in Equation(\ref{eq:4}) is presented in Equation (\ref{eq:6}), which can be easily implemented by use of a two-input look-up table. For $d-1$ operands, corresponding to a degree-d check node, the “box-plus” operation can be computed via repeated computation of $\beta$ function as in Equation (\ref{eq:7}).
\begin{equation}\label{eq:3}
\left| {L_{ij}^t} \right| = {\phi ^{ - 1}}\left( {\sum\limits_{j' \in \Psi (i)/j} {\phi (\left| {L_{j'i}^t} \right|)} } \right)
\end{equation}
\begin{equation}\label{eq:4}
\left| {L_{ij}^t} \right| = \mathop{\boxplus}_{j' \in \Psi (i)/j} {\left| {L_{j'i}^t} \right|}
\end{equation}
\begin{equation}\label{eq:5}
\phi (x) = {\phi ^{ - 1}}(x) = {\rm{ - }}\ln \left( {\tanh \left( {\frac{x}{2}} \right)} \right)
\end{equation}
\begin{equation}\label{eq:6}
{L_1} \boxplus {L_2} = \beta ({L_1},{L_2}) = 2{\tanh ^{ - 1}}\left( {\tanh \left( {\frac{{{L_1}}}{2}} \right)\tanh \left( {\frac{{{L_2}}}{2}} \right)} \right)
\end{equation}
\begin{equation}\label{eq:7}
\mathop{\boxplus}\limits_{i = 1}^{d - 1} \left| {{L_i}} \right| = \beta \left( {\left| {{L_1}} \right|,\beta \left( {\left| {{L_2}} \right|, \cdots ,\beta \left( {\left| {{L_{d - 2}}} \right|,\left| {{L_{d - 1}}} \right|} \right)} \right)} \right)
\end{equation}

In order to further simplify the computation complexity of CNs, a series of approximations are proposed including Min-Sum (MS)\cite{Decode_Fossorier1999Reduced}, Offset Min-Sum (OMS)\cite{Decode_chen2002near}, Normalized Min-Sum (NMS)\cite{Decode_chen2002near}, Approximate Min$^{*}$ Decoder\cite{Decode_Jones2004Approximate}, Richardson / Novichkov (RN)\cite{Decode_richardson2005node}, Reduced-Complexity Box-Plus (RCBP) decoder\cite{Decode_Viens2008A}, and etc. The most commonly used approximations accoding to the existing literatures are MS-based decoding algorithm including MS, OMS and NMS. Although these approximations drastically reduce the computation complexity, it is hard to obtain high reconciliation efficiency in QKD environment using these approximations. RN decoder is able to obtain a better reconciliation efficiency than OMS and NMS but it is not software-friendly. Based on the considerations as above, we adopt RCBP algorithm which can be easily implemented in CPU environment with the potential to achieve high efficiency. 

The variable node processing using layered scheduling are presented in Equations (\ref{eq:8}) and (\ref{eq:9}). $E_j'$ and $E_j$ are the soft value of VN before and after layered processing respectively. The sign of $E_j$ is the hard decision estimate of the $\rm{VN_j}$ and the magnitude means the reliability of the decision. During the iterative decoding process, the hard decision estimates and their reliabilities are updated. Indeed, in layered decoding, only the latest ${E_j}$ and ${L_{ij}}$ are stored in memory. $L_{ji}$ are computed on the fly as in Equation (\ref{eq:8}). 
\begin{equation}\label{eq:8}
L_{ji}^t = {E_j'} - L_{ij}^{t-1}
\end{equation}
\begin{equation}\label{eq:9}
{E_j} = L_{ji}^t + L_{ij}^t
\end{equation}

A stopping criterion is inherited in LDPC code to detect the correct word. At the end of each decoding iteration, parity check constraints are tested with the latest $E_j$. A correct code can be obtained only when all the parity-check constraints are satisfied. If a correct code has not been detected until the iteration number reaches the predefined maximum iteration number, the decoding process will terminate in failure. Even if all parity check constraints are satisfied, there may exists undetected errors. During decoding process, SPA decoding algorithm may converge to an inappropriate code satisfying the constraints as well. This problem rarely appears and can be solved by the subsequent processing, so called verification.

\section{Quantized RCBP-based LDPC decoder}
\label{sec:optimization}
In this section, a high throughput and efficiency LDPC decoder is proposed. High throughput performance mainly comes from quantization process which is detailed in Section \ref{sec:quantization}. Quantization process is able to improve the throughput by a factor of four, but it is certain to have negative impacts on reconciliation efficiency. In order to maintain the high efficiency as the LDPC decoder of floating-point version, the improved RCBP-based check node processing and saturation-oriented variable node processing are proposed in Section \ref{sec:3.2} and Section \ref{sec:3.3} respectively.
\subsection{Quantization}
\label{sec:quantization}
Quantization is a common used approach in hardware implementations to reduce the storage consumption, while it is not common in CPU implementations since the storage space is not a big issue in most CPU applications. Nevertheless, quantization is an effective technique to decrease processing delay when the performance of throughput is pursued in a CPU implementation. The reduction of processing delay using quantization mainly comes from the increasing of SIMD unit utilization and cache hit rates. 

In practice, SIMD processing are performed using several sets of instruction extensions supported by specific architectures. In this research, we focus on x86 architecture. Two commonly used instruction extensions are Streaming SIMD Extensions (SSE) with 128-bit registers and Advanced Vector Extensions (AVX) with 256-bit registers\cite{SIMD_deilmann2012guide}. Only 4 and 8 floating-point computations can be processed in one clock cycle in SSE and AVX mode respectively. If the data are quantized as 8-bit fixed-point numbers, 16 and 32 computations can be processed in one clock cycle in these two models respectively, which means nearly four times speed-up.

Moreover, the memory access latency varies among different levels of caches in multi-level cache architecture. It takes about 4/10/40 clock cycles to access L1/L2/L3 cached data\cite{Cache_levinthal2009performance}, and more than thousand clock cycles for uncached data. Unfortunately, the sizes of fast caches are quite limited. For instance, an i7-4790K CPU contains four cores. Each core has only one 32KB L1 data cache and one 256KB L2 cache respectively. The size of L3 cache is 8MB, but it is shared by all cores. Reducing memory size helps in fitting better into faster cache, which decreases execution time. If 8-bit fixed-point  is used instead of floating-point representation, the memory size can be reduced by at least four, leading to an improvement in throughput.

As mentioned above, the throughput can be improved by a factor of at least four via using 8-bit fixed-point representation, which is the major source of throughput improvement. However, an efficient implementation of LDPC decoder using 8-bit fixed point representation is a challenging task. The issue of quantization precision affects reconciliation efficiency negatively. To overcome the bad influence, algorithm optimizations of check node and variable processing are presented in the following sections. 

\subsection{Improved RCBP-based check node processing}
\label{sec:3.2}
In QKD environment, the reconciliation efficiency of original RCBP approximation still lags behind BP algorithm, especially for low QBERs which most DV-QKD systems focus on. The efficiency lost mainly comes from the rough quantization. Such practical issue motivates us to propose an improved RCBP approximation applying more accurate quantization step size. Meanwhile, to maintain the same range of message representation as in original RCBP approximation, a larger scale look-up table is designed. Let $\tau (p,q)$ represent the integer output value of the loop-up table. Thus, $\tau (p,q)$ is the integer representation of the function value $\beta \left( {p\Delta  + \frac{\Delta }{2},q\Delta  + \frac{\Delta }{2}} \right)$.

It is noticed that the pre-stored loop-up table is not always a good solution. As mentioned in Section \ref{sec:quantization}, four clock cycles are needed even if L1 cached data is accessed. If the loop-up table can be described by a simple equation, the calculation time may by significantly less than the access time of the look-up table, especially when SIMD technique is applied. Thus, according to the process described in the literature\cite{Decode_Viens2008A}, a new look-up table generation algorithm is designed as Algorithm \ref{algorithm:1}, involving a series of simple operations. The result of the comparison $(d<c)$ returns 1 if true and 0 if false. The constant number $\rm{{MSG}_{max}}$ is determined by the size of loop-up table. For example, if the input $p$ and $q$ are 6-bit input,  $\rm{{MSG}_{max}}$ is set as 63.
\begin{algorithm}[ht]
	\caption{Computation of $\tau (p,q)$}\label{algorithm:1}
	\KwIn{Quantified intger $p, q$}
	\KwOut{$\tau (p,q)$}
	$a \leftarrow MIN(p, \rm{{MSG}_{max}}) $\;
	$b \leftarrow MIN(q, \rm{{MSG}_{max}}) $\;
	$g \leftarrow MAX(a, b) $\;
	$l \leftarrow MIN(a, b) $\;
	$d \leftarrow g-l $\;
	\If{$l > 2$}
		{$\tau \leftarrow l-(d<2)-(d<6)$\;}
	\Else
		{$\tau \leftarrow MAX(l-(d<4),0)$\;}
\end{algorithm}
\subsection{Saturation-oriented variable node processing}
\label{sec:3.3}
Variable node processing is simple, since only addition operations are involved. However, some novel operations have to be designed to avoid the efficiency lost caused by the saturation of quantized values. For instance, during the decoding process, the reliability of some variable nodes may be very high. Assuming that the integer representation of a variable node's soft value is 330 and the increment values of the adjacent check nodes computed by equation $L_{ij}^t - L_{ij}^{t-1}$ are -50, -40 and -70, respectively. At the end of the decoding iteration, the soft value of the variable node should be updated to 170. However, using 8-bit fixed-point representation, the original and updated soft value are 127 and -33 respectively. Not only the magnitude but also the sign is incorrect. This sort of situation may lead to an incorrect result, which should be avoided.

To overcome such issue, a variable node updating rule applying in layered scheduling is proposed in Algorithm \ref{algorithm:2}. The updating rule is simple that only the variable nodes whose magnitudes have not reach maximum should be updated. The function $ABS(x)$ returns absolute value of the input $x$. The constant number $\rm{E_{max}}$ is the predefined maximum magnitude and $\rm{{ITER}_{max}}$ represents the maximum number of iterations. Using 8-bit fixed-point representation, the value of $\rm{E_{max}}$ is set as 127. The variable $t$ indicates the iteration number. 
\begin{algorithm}[ht]
	\caption{Saturation-oriented variable node processing}\label{algorithm:2}
	\KwIn{$L_{ji}^t$, $L_{ij}^t$}
	\KwOut{$E_j$}
	\For{$t=1 \to \rm{{ITER}_{max}}$}
		{\ForEach{$i \in C$}
			{\ForEach{$j \in \Psi(i)$}
				{\If{$ABS(E_j) \ne \rm{E_{max}}$}
					{${E_j} \leftarrow L_{ji}^t + L_{ij}^t$\;}}}}
\end{algorithm}

The proposed updating rule is effective in most cases. Nevertheless, this rule may introduce a new problem: an incorrect hard decision of a variable node with the maximal magnitude will no longer update. Although the decoding process will terminate in failure under that situation, the probability of such situation is so small that it has little effect on reconciliation performance. 

\section{Experiments and results}
\label{sec:experiment}
\subsection{Experiment environment}
\label{environment}
Structured QC-LDPC codes are used in the experiment\cite{ryan2009channel}. A QC-LDPC code is defined by a base matrix of size $M_b \times N_b$. Each element of the base matrix is a sub-matrix with the expansion-factor $Z$. Each nonzero entry is replaced by a cyclically shifted identity matrix while each zero entry is replaced by a all-zero matrix. To evaluate the performance of the reconciliation module, a set of QC-LDPC codes are constructed with frame length of 100kb using the finite field approach\cite{QC_lan2007construction}. This approach ensures that the girth of a LDPC code is at least 6. The check and variable node degree distributions are both irregular and optimized using the Density Evolution (DE) algorithm\cite{DE_elkouss2009efficient}. The masking matrices are constructed by using standard PEG algorithm\cite{PEG_Hu2005Regular} with the obtained degree distributions. It is noticed that the structural properties of QC-LDPC codes contribute to the obtaining of good degree distributions. 

The rate-adaptive LDPC reconciliation protocol using bi-directional and interactive approach is applied in the implementation. The positions of shortened and punctured bits are chosen based on the following principles. The positions with low column weights are chosen as puncturing bits first. The shortened bits are selected in turn from the puncturing bits when addition information revealing is necessary. 

For software implementations of LDPC reconciliation in DV-QKD systems, the optimal combination of throughput and efficiency is presented in Ref\cite{Dixon2014High}. The evaluation platform of our experiment and Ref\cite{Dixon2014High} are detailed in Table \ref{tab:1}. It can be seen from the table that the CPU platforms of i7-6700HQ and X5675 are used in our experiment and Ref\cite{Dixon2014High} respectively. 
In general, the performance of CPU platforms are evaluated by the core number and working frequency of the processor. The base frequency of i7-6700HQ is only 2.6GHz which is lower than X5675. Thanks to the more advanced turbo boost technology, the max turbo frequency of i7-6700HQ is able to reach 3.50GHz which is almost the same as X5675. However, the max turbo frequency is achieved only when a single processor core is used. When all processor cores are switched on, the working frequencies of both i7-6700HQ and X5675 will reduce to about 3.1GHz. Besides, the core number of i7-6700HQ and X5675 are four and six respectively. Thus, comprehensively considering the effects of core number and working frequency, the performance of our CPU platform is only two thirds of that used in Ref\cite{Dixon2014High}. In addition, the GPU platform that is more powerful than i7-6700HQ and X5675 is not adopted in our experiment, but is used in Ref\cite{Dixon2014High}. 

\begin{table}[ht]
	\caption{Evaluation platforms}
	\label{tab:1}       
	\centering
	\begin{tabular}{cccc}
		\hline\noalign{\smallskip}
		  & Ours & CPU\cite{Dixon2014High} & GPU\cite{Dixon2014High}\\
		\noalign{\smallskip}\hline\noalign{\smallskip}
		Processor & Intel i7-6700HQ & Intel X5675 & NVidia M2090 \\
		Number of Cores & 4 & 6 & 512\\
		Vertical Segment & Mobile & Server & Server\\
		Base Frequency & 2.60GHz & 3.06GHz & 1.3GHz \\
		Max Turbo Frequency & 3.50GHz & 3.46GHz & --- \\
		\noalign{\smallskip}\hline
	\end{tabular}
\end{table}

\subsection{Experiment results}
\label{sec:4.2}
\begin{figure}[ht]
	\centering
	\includegraphics[width=0.75\textwidth]{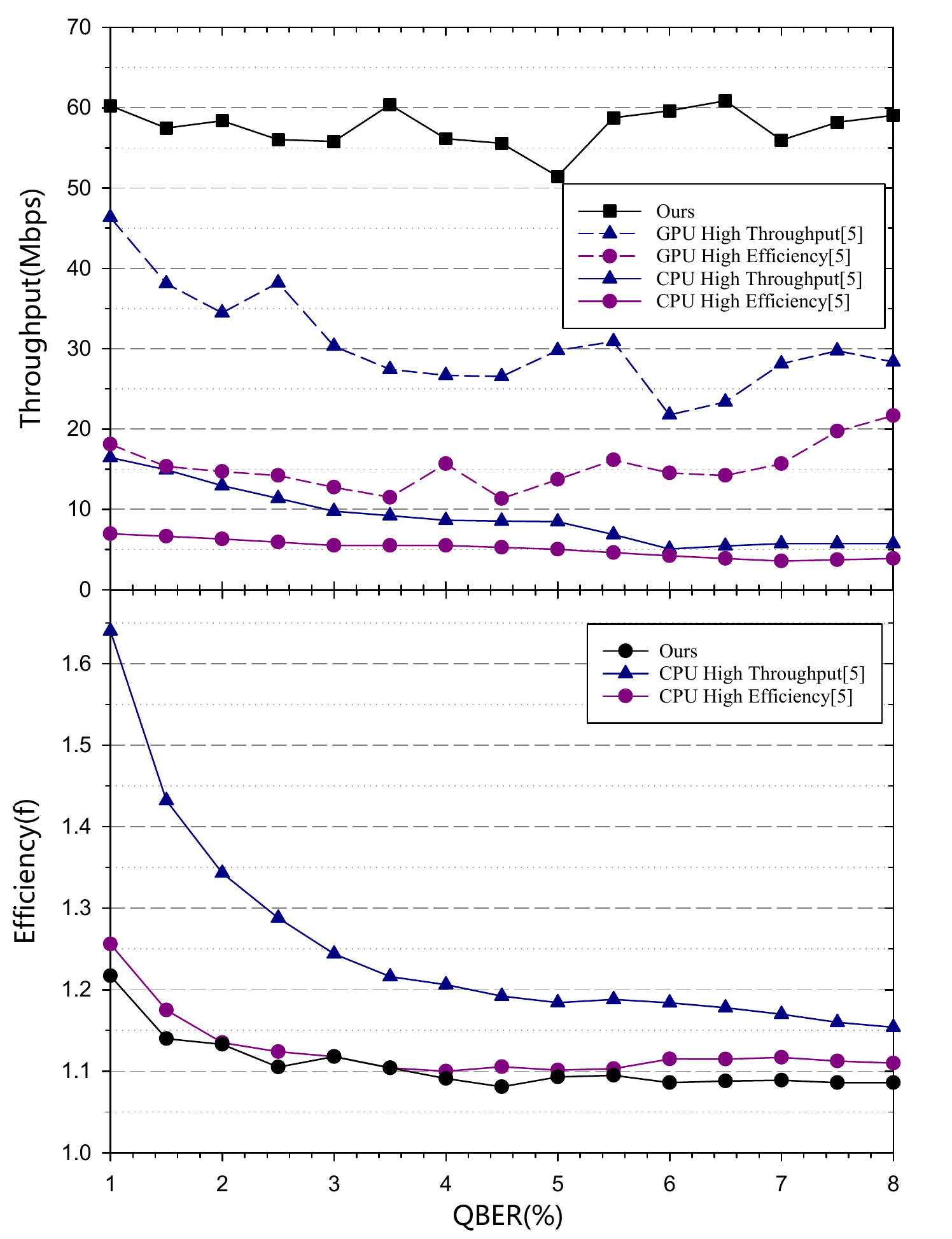}
	\caption{Throughput(upper panel) and efficiency(lower panel). It is noticed that efficiencies of CPU and GPU are same.}
	\label{fig:experiment1}
\end{figure}

Simulation results are presented in Figure \ref{fig:experiment1}. It can be seen that the throughput and efficiency surpass comparative schemes during the whole range of QBER. The efficiency and throughput are achievable when the number of errors in reconciliation data blocks are known. In the implementation, SIMD and multi-core techniques are applied to intra-frame and inter-frame parallelism respectively. 

The reconciliations modes called high throughput and high efficiency are provided in Ref\cite{Dixon2014High}. The high throughput performance is achieved by decreasing  the reconciliation efficiency of high efficiency mode properly, leading to a faster convergence speed. Using the same approach, the throughput of our implementation can further improve as well. Thus, only the high efficiency mode with the same level of efficiency is concentrated on. Compared to CPU high efficiency mode in Ref\cite{Dixon2014High}, an average speedup factor of $\times$11.3 is obtained. The high throughput improvement mainly comes from quantization and simplified check-node processing which increase the throughput at least four times. Taking advantage of  AVX-2 instruction extensions can double the throughput as well. Other implementation optimizations, such as early termination and parallel scheme optimization, also contribute to the throughput improvement. Even though compared with the GPU implementation, the average speedup factor still reaches $\times$3.8. The throughput improvement is gained by the better balance among multiple influence factors of throughput. From the angle of computational resources, GPU device is undoubtedly the most powerful one among the three evaluation platforms in Table \ref{tab:1}. However, the throughput of a practical application is not only determined by the amount of computational resources, but also affected by multiple factors. For instance, Wang et al. \cite{Wang2011A} has demonstrated
that the bottleneck of the LDPC decoder on GPU is the slow memory accesses. Besides, layered scheduling that can reduce the decoding latency by a factor of two is used in our implementation. But due to the data dependencies between consecutive layers, it is not suitable for GPU implementations. 

As depicted in Figure \ref{fig:experiment1}, not only the throughput but also the efficiency surpasses comparative schemes. The efficiency improvement is achieved by the use of algorithm optimization in Section \ref{sec:optimization} and the interactive rate-adaptive LDPC reconciliation protocol. In the original rate-adaptive reconciliation protocol \cite{LDPC_David2011Information}, interaction is only an optional step. However,  multi-interaction is an essential step in our implementation to further improve reconciliation efficiency. Although the interactions have negative impacts on throughout performance, it is worthwhile from a comprehensive point of view.

In addition to the comparison with software implementations, a performance comparison with the latest hardware(FPGA) implementation\cite{QKDSYSTEM_Z201810} is also made in Table \ref{tab:2}. An average throughput of 55Mbps with efficiency about 1.16 has been achieved in Ref \cite{QKDSYSTEM_Z201810}. To our best knowledge, this work is the best hardware LDPC reconciliation in DV-QKD systems till now. The throughput of our implementation does not apparently outperform that in Ref \cite{QKDSYSTEM_Z201810}. However, the throughput of our implementation will be further increased by using more powerful CPU processor. As can be seen from Table \ref{tab:2}, the throughput has reached 122.17Mbps by using the latest i9-9900K processor, which is more than two times faster than comparative implementations. 

\begin{table}[ht]
	\caption{Performance Comparison}
	\label{tab:2}       
	\centering
	\begin{tabular}{ccccc}
		\hline\noalign{\smallskip}
		Refs. & Target & Device & Throughput(Mbps)  & Efficiency \\ 
		\noalign{\smallskip}\hline\noalign{\smallskip}
		This work & CPU & Intel i9-9900K & 122.17 & 1.108 \\
		This work & CPU & Intel i7-4790K & 65.59  & 1.108 \\
		This work & CPU & Intel i7-6700HQ & 57.60 & 1.108 \\
		\cite{Dixon2014High} & GPU & NVidia M2090 & 30.70 & 1.250\\
		\cite{Dixon2014High} & CPU & Intel X5675 & 9.00 & 1.250\\
		\cite{QKDSYSTEM_Z201810} & FPGA & Altera Stratix V 5SGXA7 & 55 & $\approx$ 1.16 \\
		\noalign{\smallskip}\hline
	\end{tabular}
\end{table}

\begin{figure}[ht]
	\centering
	\includegraphics[width=1\textwidth]{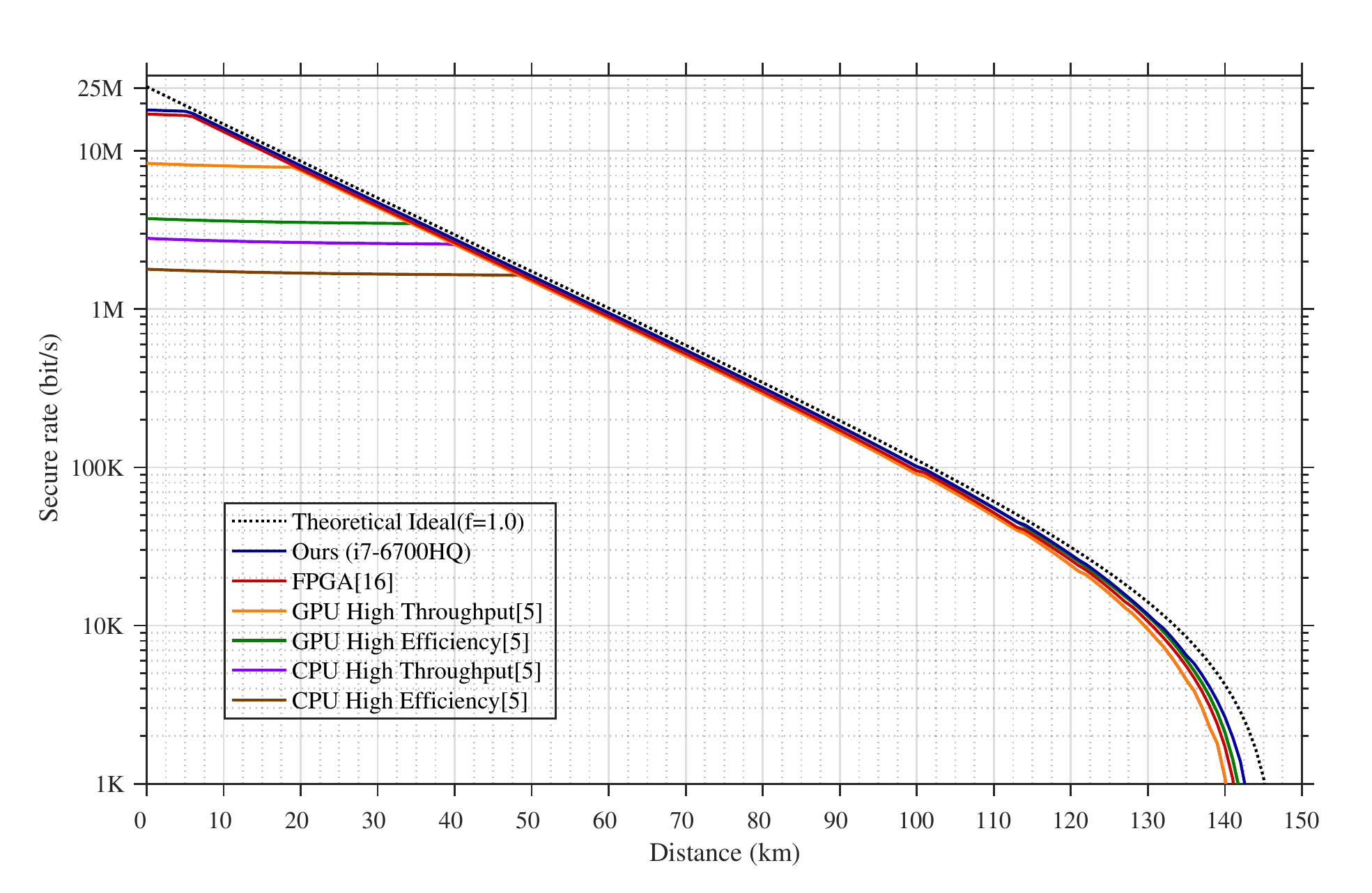}
	\caption{Secure key rate calculated for a typical high speed QKD system using real  implementations}
	\label{fig:experiment2}
\end{figure}

Reconciliation module is not a standalone system, but a part of entire QKD system. The important metric of a QKD system is the final secure key rate. Thus, using the parameter settings in Ref\cite{QKDSYSTEM_Z201810} which is the representative of state of the art DV-QKD systems, the secure rate of the QKD system is simulated. The secure key rates at different distances by using ideal reconciliation are calculated as a benchmark. The ideal reconciliation is assumed to have unlimited throughput and disclose the theoretical minimum amount of information leakage. Fig \ref{fig:experiment2} shows the secure key rates calculated for the high speed QKD system using both ideal and theoretical reconciliation modules. At short distances, the processing demands of reconciliation module is large. Once the throughput of reconciliation is not sufficient, the final secure key rate will be fixed. As the distance becomes longer, the throughput is to satisfy the processing demand. In such case, the secure key rate is determined principally by the reconciliation efficiency. Above all, both throughput and efficiency are important to final secure key rates of high speed QKD systems. Because both the throughput and efficiency of our reconciliation module are higher than those of comparative ones, the final secure key rates are higher as well.  

What's more, our implementation is applicable to nearly all kinds of DV-QKD systems. For a low speed QKD system, low performance CPU is sufficient to reduce the cost of whole system. Moreover, blind reconciliation with more communication rounds can be applied to achieve higher efficiency since the throughput of our implementation is much more than demand. For a high speed DV-QKD system, the additional GPU hardware is no longer needed when our scheme is applied. 

\section{Conclusion}
\label{sec:conclusion}
In this research, a high throughput and efficiency LDPC reconciliation scheme on a low cost platform is proposed, which is applicable to both low speed and high speed QKD systems. The proposed scheme is adaptive to different QBERs ranging from 1\% to 8\% with maximum throughput up to 60Mbps. Except for the high performance, our scheme has a good extendibility. First, although this is a software scheme designed for CPU, the optimization is suitable for hardware implementation as well. Secondly, by means of adjusting the quantization step size and the size of loop-up table, the decoding algorithm may also fit for CV-QKD systems. It is noticed that AVX-512 instruction sets are now available in some high-end CPUs, which are able to increase the throughput of the proposed scheme by up to a factor of two. Thus, the throughput of our implementation has the potential to further increase, meeting the requirements of faster QKD systems or short distance applications in the future. 

\section{Acknowledgements}
\noindent
This work is supported by the National Natural Science Foundation of China (Grant Number: 61531003, 61771168, 61702224), Space Science and Technology Advance Research Joint Funds (6141B06110105). Many thanks are extended to Prof. Z.F. Han and Prof. T. Liu for the helpful discussion.


\begin{thebibliography}{10}
	\providecommand{\url}[1]{{#1}}
	\providecommand{\urlprefix}{URL }
	\expandafter\ifx\csname urlstyle\endcsname\relax
	\providecommand{\doi}[1]{DOI~\discretionary{}{}{}#1}\else
	\providecommand{\doi}{DOI~\discretionary{}{}{}\begingroup
		\urlstyle{rm}\Url}\fi
	
	\bibitem{Bennett2014Quantum}
	Bennett, C.H., Brassard, G.: Quantum cryptography: public key distribution and
	coin tossing.
	\newblock Theoretical Computer Science \textbf{560}, 7--11 (2014)
	
	\bibitem{Gisin2001Quantum}
	Gisin, N., Ribordy, G., Tittel, W., Zbinden, H.: Quantum cryptography.
	\newblock Reviews of Modern Physics \textbf{74}(1), 145--195 (2001)
	
	\bibitem{RENATO2008SECURITY}
	Renner, R.: Security of quantum key distribution.
	\newblock International Journal of Quantum Information \textbf{6}(1), 1--127
	(2008)
	
	\bibitem{Walenta2014A}
	Walenta, N., Burg, A., Caselunghe, D., Constantin, J., Gisin, N., Guinnard, O.,
	Houlmann, R., Junod, P., Korzh, B., Kulesza, N.: A fast and versatile quantum
	key distribution system with hardware key distillation and wavelength
	multiplexing.
	\newblock New Journal of Physics \textbf{16}(1), 83--97 (2014)
	
	\bibitem{Dixon2014High}
	Dixon, A.R., Sato, H.: High speed and adaptable error correction for megabit/s
	rate quantum key distribution.
	\newblock Scientific Reports \textbf{4}, 7275 (2014)
	
	\bibitem{Li2014Study}
	Li, Q., Le, D., Mao, H., Niu, X., Liu, T., Guo, H.: Study on error
	reconciliation in quantum key distribution.
	\newblock Quantum Information $\&$ Computation \textbf{14}(13-14), 1117--1135
	(2014)
	
	\bibitem{Cascade_brassard1993secret}
	Brassard, G., Salvail, L.: Secret-key reconciliation by public discussion.
	\newblock In: Workshop on the Theory and Application of Cryptographic
	Techniques, pp. 410--423. Springer (1993)
	
	\bibitem{Cascade_sugimoto2000study}
	Sugimoto, T., Yamazaki, K.: A study on secret key reconciliation protocol.
	\newblock IEICE Transactions on Fundamentals of Electronics, Communications and
	Computer Sciences \textbf{83}(10), 1987--1991 (2000)
	
	\bibitem{Cascade_Nakassis2004Expeditious}
	Nakassis, A., Bienfang, J.C., Williams, C.J.: Expeditious reconciliation for
	practical quantum key distribution.
	\newblock In: Proceedings of SPIE - The International Society for Optical
	Engineering, vol. 5436, pp. 28--35. International Society for Optics and
	Photonics (2004)
	
	\bibitem{Cascade_Yan2008Information}
	Yan, H., Ren, T., Peng, X., Lin, X., Jiang, W., Liu, T., Guo, H.: Information
	reconciliation protocol in quantum key distribution system.
	\newblock In: Natural Computation, 2008. ICNC'08. Fourth International
	Conference on, vol.~3, pp. 637--641. IEEE (2008)
	
	\bibitem{Cascade_Pedersen2013High}
	Pedersen, T.B., Toyran, M.: High performance information reconciliation for qkd
	with cascade.
	\newblock Quantum Information $\&$ Computation \textbf{15}(5-6), 419--434
	(2013)
	
	\bibitem{Cascade_Pacher2015An}
	Pacher, C., Grabenweger, P., Martinez-Mateo, J., Martin, V.: An information
	reconciliation protocol for secret-key agreement with small leakage.
	\newblock In: IEEE International Symposium on Information Theory, pp. 730--734.
	IEEE (2015)
	
	\bibitem{Polar_Arikan2008Channel}
	Arikan, E.: Channel polarization: A method for constructing capacity-achieving
	codes for symmetric binary-input memoryless channels.
	\newblock IEEE Transactions on Information Theory \textbf{55}(7), 3051--3073
	(2009)
	
	\bibitem{Polar_jouguet2014high}
	Jouguet, P., Kunz-Jacques, S.: High performance error correction for quantum
	key distribution using polar codes.
	\newblock Quantum Information $\&$ Computation \textbf{14}(3-4), 329--338
	(2014)
	
	\bibitem{Polar_Yan2018An}
	Yan, S., Wang, J., Fang, J., Lin, J., Wang, X.: An improved polar codes-based
	key reconciliation for practical quantum key distribution.
	\newblock Chinese Journal of Electronics \textbf{27}(2), 250--255 (2018)
	
	\bibitem{QKDSYSTEM_Z201810}
	Yuan, Z., Plews, A., Takahashi, R., Doi, K., Tam, W., Sharpe, A.W., Dixon,
	A.R., Lavelle, E., Dynes, J.F., Murakami, A.: 10-mb/s quantum key
	distribution.
	\newblock Journal of Lightwave Technology \textbf{36}(16), 3427--3433 (2018)
	
	\bibitem{LDPC_David2011Information}
	Elkouss, D., MartinezMateo, J., Martin, V.: Information reconciliation for
	quantum key distribution.
	\newblock Quantum Information $\&$ Computation \textbf{11}(3), 226--238 (2011)
	
	\bibitem{LDPC_Martinez2012Blind}
	Martinez-Mateo, J., Elkouss, D., Martin, V.: Blind reconciliation.
	\newblock Quantum Information $\&$ Computation \textbf{12}(9-10), 791--812
	(2012)
	
	\bibitem{LDPC_Kiktenko2017Symmetric}
	Kiktenko, E., Truschechkin, A., Lim, C., Kurochkin, Y., Federov, A.: Symmetric
	blind information reconciliation for quantum key distribution.
	\newblock Physical Review Applied \textbf{8}(4), 044017 (2017)
	
	\bibitem{GPU_Wang2018High}
	Wang, X., Zhang, Y., Yu, S., Guo, H.: High speed error correction for
	continuous-variable quantum key distribution with multi-edge type LDPC code.
	\newblock Scientific reports \textbf{8}(1), 10543 (2018)
	
	\bibitem{GPU_Milicevic2018Quasi}
	Milicevic, M., Chen, F., Zhang, L.M., Gulak, P.G.: Quasi-cyclic multi-edge ldpc
	codes for long-distance quantum cryptography.
	\newblock NPJ Quantum Information \textbf{4}(1), 1--9 (2018)
	
	\bibitem{CPU_Gal2016High}
	Gal, B.L., Jego, C.: High-throughput multi-core LDPC decoders based on x86
	processor.
	\newblock IEEE Transactions on Parallel $\&$ Distributed Systems
	\textbf{27}(5), 1373--1386 (2016)
	
	\bibitem{LDPC_gallager1962low}
	Gallager, R.: Low-density parity-check codes.
	\newblock IRE Transactions on information theory \textbf{8}(1), 21--28 (1962)
	
	\bibitem{ryan2009channel}
	Ryan, W., Lin, S.: Channel codes: classical and modern.
	\newblock Cambridge University Press (2009)
	
	\bibitem{LDPC_Mackay1999Good}
	MacKay, D.J.: Good error-correcting codes based on very sparse matrices.
	\newblock IEEE Transactions on Information Theory \textbf{45}(2), 399--431
	(1999)
	
	\bibitem{LDPC_Hocevar2004}
	Hocevar, D.E.: A reduced complexity decoder architecture via layered decoding
	of LDPC codes.
	\newblock In: IEEE Workshop on Signal Processing Systems, pp. 107--112. IEEE
	(2004)
	
	\bibitem{Decode_Jones2004Approximate}
	Jones, C., Vall{\'e}s, E., Smith, M., Villasenor, J.: Approximate-min
	constraint node updating for LDPC code decoding.
	\newblock In: Military Communications Conference, 2003. MILCOM'03. 2003 IEEE,
	vol.~1, pp. 157--162. IEEE (2003)
	
	\bibitem{Jones2007Functions}
	Jones, C., Dolinar, S., Andrews, K., Divsalar, D., Zhang, Y., Ryan, W.:
	Functions and architectures for LDPC decoding.
	\newblock In: IEEE Information Theory Workshop, pp. 577--583. IEEE (2007)
	
	\bibitem{Decode_Fossorier1999Reduced}
	Fossorier, M.P., Mihaljevic, M., Imai, H.: Reduced complexity iterative
	decoding of low-density parity check codes based on belief propagation.
	\newblock IEEE Transactions on communications \textbf{47}(5), 673--680 (1999)
	
	\bibitem{Decode_chen2002near}
	Chen, J., Fossorier, M.P.: Near optimum universal belief propagation based
	decoding of low-density parity check codes.
	\newblock IEEE Transactions on Communications \textbf{50}(3), 406--414 (2002)
	
	\bibitem{Decode_richardson2005node}
	Richardson, T., Novichkov, V.: Node processors for use in parity check decoders
	(2005).
	\newblock US Patent 6,938,196
	
	\bibitem{Decode_Viens2008A}
	Viens, M., Ryan, W.E.: A reduced-complexity box-plus decoder for LDPC codes.
	\newblock In: International Symposium on Turbo Codes and Related Topics, pp.
	151--156. IEEE (2008)
	
	\bibitem{SIMD_deilmann2012guide}
	Deilmann, M., et~al.: A guide to vectorization with intel c++ compilers.
	\newblock Intel Corporation, April  (2012)
	
	\bibitem{Cache_levinthal2009performance}
	Levinthal, D.: Performance analysis guide for intel core i7 processor and intel
	xeon 5500 processors.
	\newblock Intel Performance Analysis Guide \textbf{30}, 18 (2009)
	
	\bibitem{QC_lan2007construction}
	Lan, L., Zeng, L., Tai, Y.Y., Chen, L., Lin, S., Abdel-Ghaffar, K.:
	Construction of quasi-cyclic LDPC codes for AWGN and binary erasure channels:
	A finite field approach.
	\newblock IEEE Transactions on Information Theory \textbf{53}(7), 2429--2458
	(2007)
	
	\bibitem{DE_elkouss2009efficient}
	Elkouss, D., Leverrier, A., All{\'e}aume, R., Boutros, J.: Efficient
	reconciliation protocol for discrete-variable quantum key distribution.
	\newblock In: IEEE International Conference on Symposium on Information Theory,
	pp. 1879--1883. IEEE (2009)
	
	\bibitem{PEG_Hu2005Regular}
	Hu, X.Y., Eleftheriou, E., Arnold, D.M.: Regular and irregular progressive
	edge-growth tanner graphs.
	\newblock IEEE Transactions on Information Theory \textbf{51}(1), 386--398
	(2005)
	
	\bibitem{Wang2011A}
	Wang, G., Wu, M., Yang, S., Cavallaro, J.R.: A massively parallel
	implementation of qc-ldpc decoder on gpu.
	\newblock In: Application Specific Processors (2011)
	
	
\end{thebibliography}

\end{document}